\documentclass[
  journal=pasa,
  manuscript=research-paper,
  year=2020,
  volume=37,
]{cup-journal}
\usepackage[plainpages=false, colorlinks=true, anchorcolor=blue, linkcolor=blue, citecolor=blue, bookmarks=false]{hyperref}
\usepackage{amsmath}
\usepackage[nopatch]{microtype}
\usepackage{booktabs}
\usepackage{orcidlink}
\newcommand{\bthis}[1]{\textcolor{black}{#1}}
\title{Low-frequency pulse-jitter measurement with the uGMRT I : PSR~J0437$-$4715}

\author{Tomonosuke Kikunaga\,\orcidlink{0000-0002-5016-3567}}
\affiliation{Kumamoto University, Graduate School of Science and Technology, Kumamoto, 869-8555, Japan}
\alsoaffiliation{Australia Telescope National Facility, CSIRO Space and Astronomy, P.O. Box 76, Epping, NSW 1710, Australia}
\email[T. Kikunaga]{amqmysuto@gmail.com}

\author{Shinnosuke Hisano\,\orcidlink{0000-0002-7700-3379}}
\affiliation{International Research Organization for Advanced Science and Technology, Kumamoto University, 2-39-1 Kurokami, Kumamoto 860-8555, Japan}

\author{Neelam Dhanda Batra\,\orcidlink{0000-0003-0266-0195}}
\affiliation{Department of Physics and Astrophysics, Delhi University, Delhi 110007, India}

\author{Shantanu Desai\,\orcidlink{0000-0002-0466-3288}}
\affiliation{Department of Physics, IIT Hyderabad, Kandi, Telangana 502284, India}

\author{Bhal Chandra Joshi\,\orcidlink{0000-0002-0863-7781}}
\affiliation{National Centre for Radio Astrophysics (TIFR), Pune, India}
\alsoaffiliation{Department of Physics, Indian Institute of Technology, Roorkee, India}

\author{Manjari Bagchi\,\orcidlink{0000-0001-8640-8186}}
\affiliation{The Institute of Mathematical Sciences, C. I. T. Campus, Taramani, Chennai 600113, India}
\alsoaffiliation{Homi Bhabha National Institute, Training School Complex, Anushakti Nagar, Mumbai 400094, India}

\author{T. Prabu \,\orcidlink{0000-0003-4038-8065}}
\affiliation{Raman Research Institute, Bangalore.}

\author{Keitaro Takahashi\,\orcidlink{0000-0002-3034-5769}}
\affiliation{Faculty of Advanced Science and Technology, Kumamoto University, 2-39-1 Kurokami, Kumamoto 860-8555, Japan}
\alsoaffiliation{International Research Organization for Advanced Science and Technology, Kumamoto University, 2-39-1 Kurokami, Kumamoto 860-8555, Japan}

\author{Swetha Arumugam\,\orcidlink{0009-0001-3587-6622}}
\affiliation{Department of Electrical Engineering, IIT Hyderabad, Kandi, Telangana 502284, India}

\author{Adarsh Bathula \,\orcidlink{0000-0001-7947-6703}}
\affiliation{Department of Physical Sciences, IISER Mohali, Punjab, India}

\author{Subhajit Dandapat \,\orcidlink{0000-0003-4965-9220}}
\affiliation{Department of Astronomy and Astrophysics, Tata Institute of Fundamental Research, Homi Bhabha Road, Navy Nagar, Colaba, Mumbai 400005, India}

\author{Debabrata Deb\,\orcidlink{0000-0003-4067-5283}}
\affiliation{The Institute of Mathematical Sciences, C. I. T. Campus, Taramani, Chennai 600113, India}

\author{Churchil Dwivedi\,\orcidlink{0000-0002-8804-650X}}
\affiliation{Department of Earth and Space Sciences, Indian Institute of Space Science and Technology, Valiamala P.O., Thiruvananthapuram 695547, Kerala, India}

\author{Yashwant Gupta\,\orcidlink{0000-0001-5765-0619}}
\affiliation{National Centre for Radio Astrophysics, Pune University Campus, Pune 411007, India}

\author{Shebin Jose Jacob\,\orcidlink{0009-0008-4208-2901}}
\affiliation{Department of Physics, Government Brennen College, Thalassery, Kannur University, Kannur 670106, Kerala, India}

\author{Fazal Kareem\,\orcidlink{0000-0003-2444-838X}}
\affiliation{Department of Physical Sciences, IISER Kolkata, West Bengal, India}
\alsoaffiliation{Center of Excellence in Space Sciences India, IISER Kolkata, West Bengal, India}

\author{Nobleson K\,\orcidlink{0000-0003-2715-4504}}
\affiliation{International Research Organization for Advanced Science and Technology, Kumamoto University, 2-39-1 Kurokami, Kumamoto 860-8555, Japan}

\author{Pragna Mamidipaka \,\orcidlink{0000-0001-9006-3192}}
\affiliation{Department of Electrical Engineering, IIT Hyderabad, Kandi, Telangana 502284, India}

\author{Avinash Kumar Paladi\,\orcidlink{0000-0002-8651-9510}}
\affiliation{Joint Astronomy Programme, Department of Physics, Indian Institute of Science, Bengaluru, Karnataka, 560012, India}

\author{Arul Pandian B\,\orcidlink{0000-0002-0417-6308}}
\affiliation{Raman Research Institute, Bangalore 560080, India}
\alsoaffiliation{Christ University, Bangalore 560029, India}

\author{Prerna Rana\,\orcidlink{0000-0001-6184-5195}}
\affiliation{Department of Astronomy and Astrophysics, Tata Institute of Fundamental Research, Homi Bhabha Road, Navy Nagar, Colaba, Mumbai 400005, India}

\author{Jaikhomba Singha\,\orcidlink{0000-0002-1636-9414}}
\affiliation{High Energy Physics, Cosmology \& Astrophysics Theory (HEPCAT) Group, 
Department of Mathematics and Applied Mathematics, University of Cape Town, 
Cape Town 7700, South Africa}

\author{Aman Srivastava\,\orcidlink{0000-0003-3531-7887}}
\affiliation{Department of Physics, IIT Hyderabad, Kandi, Telangana 502284, India}

\author{Mayuresh Surnis\,\orcidlink{0000-0002-9507-6985}}
\affiliation{Department of Physics, Indian Institute of Science Education and Research Bhopal, Bhopal Bypass Road, Bhauri, Bhopal 462 066, Madhya Pradesh, India}

\author{Pratik Tarafdar\,\orcidlink{0000-0001-6921-4195}}
\affiliation{The Institute of Mathematical Sciences, C. I. T. Campus, Taramani, Chennai 600113, India}


\begin{document}

\begin{abstract}
High-precision pulsar timing observations are limited in their accuracy by the  jitter noise that appears in the arrival time of pulses. Therefore, it is important to systematically characterise the amplitude of the  jitter noise and its variation with frequency. In this paper,  we provide jitter measurements from low-frequency wideband observations of PSR~J0437$-$4715 using data obtained as part of the Indian Pulsar Timing Array experiment. We were able to detect jitter in both the 300 - 500 MHz and 1260 - 1460 MHz observations of the upgraded Giant Metrewave Radio Telescope (uGMRT). The former is  the first jitter measurement for this pulsar below 700 MHz, and the latter is in good agreement with results from previous studies. In addition, at 300 - 500 MHz, we investigated the frequency dependence of the jitter by calculating the jitter for each sub-banded arrival time of pulses. We found that the jitter amplitude increases  with  frequency. This trend is opposite as  compared to  previous studies, indicating that there is a turnover at  intermediate frequencies. It will be possible to investigate this in more detail with uGMRT observations at 550 - 750 MHz and future high sensitive wideband observations  from next generation telescopes, such as the Square Kilometre Array. We also explored the effect of jitter on  the high precision dispersion measure (DM) measurements derived from short duration observations. We find that even though the DM precision will be better at lower frequencies due to the smaller amplitude of jitter noise, it will limit the  DM precision for high signal-to-noise observations, which are of short durations. This limitation can  be overcome by integrating for a long enough duration optimised for a given pulsar. 
\end{abstract}

\defcitealias{Parthasarathy2021}{P21}
\defcitealias{Shannon2014}{S14}

\section{Introduction}
\label{sec:introduction}

Pulsars are  rapidly rotating neutron stars that emit beamed emission, which is observed as a pulsed signal. Among the  different types of pulsars, millisecond pulsars (MSPs) have the most stable periods, making them the most precise clocks in the universe.
The precise timing of MSPs has been used to search for nanohertz gravitational waves (GWs) by multiple consortia, known as  Pulsar Timing Arrays (PTAs) \citep{Fos+1990}, such as the North American Nanohertz Observatory for Gravitational Waves \citep[NANOGrav]{McLaughlin2013}, the European Pulsar Timing Array \citep[EPTA]{Kramer2013}, the Parkes Pulsar Timing Array \citep[PPTA]{Hobbs2013,Manchester2013}, the Indian Pulsar Timing Array \citep[InPTA]{Joshi2018,Joshi2022}, the Chinese Pulsar Timing Array \citep[CPTA]{Lee2016} and the MeerTime Pulsar Timing Array  \citep[MPTA]{Bailes2020,Spiewak2022}. Recently, NANOGrav, EPTA+InPTA, PPTA, and CPTA announced evidence for nanohertz stochastic gravitational wave background (SGWB)  with statistical significance ranging between $2-4\sigma$ \citep{AgazieGWB2023,AntoniadisGWB2023,ReardonGWB2023,Xu2023}.
With the next generation, high sensitivity telescopes, such as the Square Kilometre Array (SKA) it will be possible  to obtain sufficient signal-to-noise ratio (S/N) within short observation duration to measure the pulse times of arrival (TOA).
In the future, a large number of high cadence timing observations on a much larger pulsar sample will be made during the SKA era, which will lead to significant improvements in  the sensitivity to detection of nanohertz GWs from the PTAs.
This will not only facilitate the detection of SGWB at high confidence level, but will also open our horizon towards continuous GWs from individual supermassive black hole binaries, thereby ushering in the  era of multi-wavelength GW astronomy.

In order to maximise the PTA sensitivity to GWs, a detailed characterisation of the different noise sources is of crucial importance.
There are many noise sources that contribute to the uncertainty in the TOAs.
Noise can be classified according to several characteristics, one of which is the temporal correlation.
A noise with long-term correlation is called red noise. Some examples of red noise  include  rotational irregularities in the pulsar's spin period, also known as spin noise or timing noise~\citep{Boynton1972,Cordes1980,Shannon2010}, temporal variation in the Dispersion Measure (DM) caused by the interstellar plasma \citep{Keith2013,Tarafdar22}, and errors in the Solar System ephemeris \citep{Champion2010}.
On the other hand, noise without temporal correlation is called white noise.
Another feature to classify the different sources of noise is their dependence on the observation frequency.
While dispersive delays depend on the frequency, errors in the Solar System ephemeris are achromatic.
Furthermore, whether a noise source is common to multiple pulsars or unique to each pulsar is of vital importance for distinguishing the GW signal from noise.
The PTA  experiments attempt to extract the GW signals by modeling these kinds of noise and incorporating them into their data analysis  pipelines~\citep{Chalumeau2022,Srivastava23,AgazieNoise2023,AntoniadisNoise2023,ReardonNoise2023}.
Although, many sources of noise are well understood both statistically and physically, there is  room for further improvement in noise modeling.

Jitter noise refers to  stochastic fluctuations in the pulsar timing residuals \bthis{with a white noise spectrum,} and is one of the noise sources  that should be characterised precisely  to improve pulsar timing analysis.
\bthis{We, however, note that jitter is temporally correlated in the phase of the pulsar.}
The   \bthis{white noise spectrum} also includes  a contribution from the radiometer noise originating from the instrument.  The amplitude \bthis{of the radiometer noise}  depends on the instrument and this noise can be reduced, thereby improving the S/N  ratio. 
On the other hand, jitter noise does not decrease,  even if the S/N ratio improves and its amplitude is independent of the instrument \citep{Shannon2012,Shannon2014,Lam2016}.
\bthis{Furthermore, radiometer noise is always uncorrelated in time (barring instrumental issues) and homoscedastic across the pulse phase,  while jitter is both correlated and heteroscedastic~\citep{Oslowski11}.}
Hence, this noise is considered intrinsic to each pulsar, and it occurs due to pulse shape variations on pulse-to-pulse scale.
Since it behaves like white noise, its amplitude decreases with an increase in the integration time.
However, with the shorter duration of observations, which are likely to be planned for the next generation PTA experiments due to their higher sensitivity, these experiments may suffer from jitter noise, if its implications are not understood properly. 
Therefore, it is important to investigate the nature and amplitude of the jitter noise with current observational campaigns in order to develop efficient observation strategies for the future.

For this work, we study the jitter noise in the pulsar PSR~\\J0437$-$4715. This pulsar was discovered in the Parkes 70~cm survey and is the brightest known MSP \citep{Johnston1993}.
Due to its high brightness, it has been extensively studied especially for its single-pulse behaviour.
\citep{Ables1997,Jenet1998,Vivekanand1998,Vivekanand2000,Liu2012,Osłowski2014,De2016Micro}.

This pulsar has also been the focus of previous jitter studies, which measured its jitter amplitude at different frequencies
\citep[hereafter \citetalias{Shannon2014} and \citetalias{Parthasarathy2021}, respectively]{Shannon2014,Parthasarathy2021}.
\citetalias{Shannon2014} measured its jitter amplitude using 700, 1400, and 3100\,MHz observations by the Parkes \textit{Murriyang} telescope, whereas 
\citetalias{Parthasarathy2021} investigated the wideband feature of its jitter using the MeerKAT telescope (856 - 1712 MHz).
Both studies found that the jitter amplitude of PSR~J0437$-$4715 has a weak frequency dependence, and its amplitude is larger at lower frequencies.
PSR~J0437$-$4715 has been  subsequently observed by PPTA, MPTA, and InPTA,  since it is located in the southern sky. Among these three PTAs, only InPTA covers the low frequency range (300 - 500 MHz).
Therefore, we are the only PTA, which can investigate whether the jitter amplitude of PSR~J0437$-$4715 is actually getting larger in the lower frequency bands and this investigation has important ramifications  for developing a future strategy for low-frequency timing observations.

In this paper, we report the pulse jitter measurements of PSR~J0437$-$4715 for low (300 - 500 MHz) and mid (1260 - 1460 MHz) radio frequency observations using the uGMRT. The rest of this paper is structured  as follows:
In Section \ref{sec:obs_and_dataprocessing}, we briefly describe the observations used in this work and the data processing methods used for analysis. 
In Section \ref{sec:jitter}, we describe the methodology used for estimating jitter noise and the results of these measurements.
In Section \ref{sec:discussion}, we discuss the inferences and conclusions made from our analysis. Finally, we summarize our findings and conclude in Section \ref{sec:Conclusions}.

\section{Observations and Data processing}

\label{sec:obs_and_dataprocessing}
\subsection{Observations}
\label{sec:observation} 

\begin{table}
    \caption{Band 3 observations used in this work and their estimated jitter amplitudes. The first column lists the uGMRT observation cycle. The second column shows  the date of  the observations. The third column gives the integrated S/N ratios for this band  obtained from the \texttt{pdmp} command of \texttt{PSRCHIVE} software package. The fourth column is the  observation duration in seconds. The last column shows the ECORR value scaled to one hour and these  values are plotted in Figure \ref{fig:ECORR_timeseries}. The bottom six epochs were selected to see the difference between Band 3 and 5 for  the same epochs (see  Section \ref{sec:jitteramplitude}).}
    \renewcommand{\arraystretch}{1.3}
    \centering
    \begin{tabular}{ccccc}
        \hline \hline
        Cycle & MJD & S/N & Duration (s) & ECORR (ns) \\
        \hline \hline
        41	&    59545	& 16401	  &	    1198 & 63.36 $^{+4.53}_{-4.23}$ \\
        41	&    59587	& 8667	  & 	1799 & 64.84 $^{+4.64}_{-4.33}$ \\
        41	&    59627	& 7571	  & 	1018 & 53.06 $^{+3.38}_{-3.12}$ \\
        41	&    59656	& 6602	  & 	1020 & 64.32 $^{+5.04}_{-4.49}$ \\
        41	&    59665	& 4772	  & 	1320 & 47.79 $^{+4.25}_{-3.93}$ \\
        \hline
        42	&    59692	& 5196	  & 	719  & 68.48 $^{+6.51}_{-5.64}$ \\
        42	&    59701	& 14708	  &	    720  & 52.70 $^{+5.08}_{-4.64}$ \\
        42	&    59730	& 10452	  &	    718  & 52.04 $^{+4.90}_{-4.35}$ \\
        42	&    59789	& 13307	  &	    718  & 59.32 $^{+5.78}_{-4.95}$ \\
        42	&    59800	& 11717	  &	    718  & 64.58 $^{+5.85}_{-5.16}$ \\
        \hline
        43	&    59908	& 12651	  &	    598  & 52.70 $^{+3.77}_{-3.52}$ \\
        43	&    59918	& 10150	  &	    720  & 31.76 $^{+3.06}_{-2.79}$ \\
        43	&    59928	& 13041	  &	    597  & 47.86 $^{+5.20}_{-4.51}$ \\
        43	&    59989	& 11379	  &	    660  & 53.05 $^{+5.26}_{-4.57}$ \\
        43	&    60021	& 18320	  &	    1496 & 52.98 $^{+3.43}_{-3.09}$ \\
        \hline
        44	&    60063	& 5487	  & 	600  & 52.07 $^{+6.08}_{-5.29}$ \\
        44	&    60121	& 12228	  &	    598  & 56.33 $^{+5.85}_{-5.04}$ \\
        44	&    60139	& 6250	  & 	600  & 49.49 $^{+5.56}_{-4.78}$ \\
        44	&    60160	& 13017	  &	    598  & 68.70 $^{+7.01}_{-6.04}$ \\
        44	&    60178	& 23971	  &	    600  & 50.24 $^{+5.18}_{-4.38}$ \\
        \hline
        41	&    59575	& 2596	&      1318  & 63.19 $^{+6.20}_{-5.86}$ \\
        42	&    59818	& 2627	&      719   & 53.93 $^{+6.58}_{-5.87}$ \\
        43	&    59982	& 5265	&      900   & 49.72 $^{+4.81}_{-4.25}$ \\
        43	&    60002	& 6942	&      720   & 61.90 $^{+6.40}_{-5.73}$ \\
        44	&    60055	& 4930	&      900   & 55.58 $^{+5.37}_{-4.87}$ \\
        44	&    60149	& 5211	&      720   & 53.54 $^{+5.45}_{-4.77}$ \\
        \hline
    \end{tabular}
    \label{tab:band3_observation}
\end{table}
\renewcommand{\arraystretch}{1.3}
\begin{table}[ht]
    \caption{Band 5 observations used in this work. The different columns refer to same parameters as in Table \ref{tab:band3_observation}.}
    \centering
    \begin{tabular}{ccccc}
        \hline \hline
        Cycle & MJD & S/N & Duration (s) & ECORR (ns) \\
        \hline \hline
        41	&    59545	& 1253	&  1215  &  61.92 $^{+4.43}_{-4.13}$ \\
        41	&    59565	& 1126	&  1034  &  56.47 $^{+6.89}_{-7.29}$ \\
        41	&    59575	& 1647	&  1334  &  61.92 $^{+5.97}_{-5.45}$ \\
        41	&    59627	& 1465	&  1033  &  57.15 $^{+7.24}_{-6.91}$ \\
        41	&    59665	& 1332	&  1345  &  66.51 $^{+6.94}_{-6.40}$ \\
        \hline
        42	&    59692	& 1431	&  734   &  58.28 $^{+7.01}_{-6.22}$ \\
        42	&    59782	& 938 	&  733   &  46.57 $^{+9.43}_{-9.86}$ \\
        42	&    59810	& 1044	&  733   &  49.33 $^{+8.60}_{-8.36}$ \\
        42	&    59818	& 1696	&  734   &  43.84 $^{+5.35}_{-4.77}$ \\
        42	&    59838	& 2115	&  734   &  56.47 $^{+5.45}_{-4.97}$ \\
        \hline
        43	&    59898	& 1237	&  720   &  61.92 $^{+7.56 }_{-6.73}$ \\
        43	&    59918	& 1328	&  720   &  64.90 $^{+10.15}_{-9.65}$ \\
        43	&    59959	& 730 	&  901   &  54.66 $^{+10.68}_{-11.01}$ \\
        43	&    59982	& 851 	&  901   &  61.92 $^{+7.56 }_{-6.73}$ \\
        43	&    60002	& 812 	&  721   &  42.02 $^{+12.08}_{-20.46}$ \\
        \hline
        44	&    60055	& 954 	&  901   &  60.28 $^{+8.91}_{-8.23}$\\
        44	&    60149	& 1470	&  722   &  50.93 $^{+6.76}_{-6.14}$\\
        44	&    60191	& 1450	&  720   &  52.17 $^{+6.43}_{-5.77}$\\
        \hline
    \end{tabular}
    \label{tab:band5_observation}
\end{table}

In this work, we have used the observations of PSR~J0437$-$4715 conducted using uGMRT \citep{Gupta2017} as a part of the InPTA experiment from the observation Cycle 41 to 44 (October 2021 to September 2023). 
The InPTA observations were carried out either using two sub-arrays of 10 to 15 antennae at Band 3 (300 - 500 MHz) and Band 5 (1260 - 1460 MHz), respectively or using the complete array set at Band 3.  \bthis{The voltages from individual antennas were added after
compensating for the phase difference between different
antennas to form  phased array beams for both Band 3 and
Band 5. This Nyquist sampled
time-series was converted to 1024 frequency channels
for Band 5 by taking a 2048 point discrete Fourier Transform
in the GMRT wide-band backend. The Band 3 data
were coherently dedispersed with the known DM of the
pulsar to 128 frequency channels using a real-time pipeline~\citep{De2016} and  sampled every 5.12  $\mu$s. As the dual-band observation mode with the uGMRT allows
the coherent dedispersion pipeline to be used only at
one of the bands, the spectral time-series was
recorded with 1024 channels for Band 5, in the
regular phased array mode and sampled every
40.96~$\mu$s.  Given the low DM (2.6 pc\,cm$^{-3}$) of the
pulsar, the dispersion smear across a Band 5 channel
is much smaller ($\sim$ 2 $\mu$s) than the sampling
time used for Band 5, which is approximately
equal to a phase bin across the profile. Thus,
we do not expect significant aliasing to affect
our results despite our observations being limited
by the allowed observing mode and the lack of coherent
dedispersion. } The Julian dates and S/N ratio for each epoch used for our analysis can be found in Table \ref{tab:band3_observation} and \ref{tab:band5_observation} for Band 3 and Band 5, respectively. More details about  the InPTA observations and the associated observing strategy can be found in ~\citet{Tarafdar22} and ~\citet{Joshi2022}.


\subsection{Data processing}
\label{sec:DataProcessing}
\begin{table*}
    \centering
    \caption{Parameters used in \texttt{pinta} reduction. The first column is the local oscillator frequency of the observing band. The second column is the number of phase bins. 
    The third column is the number of frequency channels. The fourth columns denotes the  observation bandwidth. The fifth column is the sampling time used for observation. The sixth column is the sideband. The seventh column is number of polarizations. The eighth column is the duration of individual sub-integrations. The last column is whether the data has been coherently dedispersed (1) or not (0). More details on the description of these parameters can be found in \citet{Susobhanan2021}.}
    \begin{tabular}{c|ccccccccc}
        \hline \hline
         Parameters & Frequency & $N_\mathrm{bins}$ & $N_\mathrm{chan}$ & Band width & $T_\mathrm{smpl}$ & Sideband & $N_\mathrm{pol}$ & $T_\mathrm{subint}$ & Coheded\\
          & [MHz] & & & [MHz] & [$\mu$s] & & & [s] & \\
         \hline
         Band 3 & 500 & 1024 & 128 & 200 & 5.12 & LSB & 1 & 10 & 1 \\
         Band 5 & 1460 & 128 & 1024 & 200 & 40.96 & LSB & 1 & 10 & 0 \\
         \hline
    \end{tabular}
    \label{tab:pipeline}
\end{table*}

The processing of  uGMRT pulsar observations  was done using \bthis{publicly available pulsar analysis tools described in this section. Here, we briefly summarize the data processing steps involved and the relevant software that was used in the analysis.
The raw spectral time-series was first reduced by converting
it into RFI mitigated partially folded profiles averaged every
10 s for each frequency channel using}
\texttt{pinta}\footnote{\url{https://github.com/inpta/pinta}}~\citep{Susobhanan2021} pipeline.  The steps involved are as follows:
\begin{enumerate}
\item  \bthis{Convert the raw spectral data to filterbank format after removing impulsive and periodic RFI using algorithms
developed for the uGMRT data with   \texttt{RFIClean}\footnote{\url{https://github.com/ymaan4/RFIClean}} \citep{Maan2021}, which automatically remove periodic RFI in the Fourier domain.}
\item \bthis{The RFI-mitigated filterbank data  were then folded
for every frequency channel using an updated ephemeris of the
pulsar every 10 s.   This was  done in the pipeline using  \texttt{DSPSR}\footnote{\url{https://dspsr.sourceforge.net}} \citep{vanStraten2011}.
Each profile folded every 10~s represents
a sub-integration in the observations consisting of all frequency
channels that are recorded (128 and 1024 in Band 3 and 5, respectively).
These partially folded profiles were then written in the standard
PSRFITS format for subsequent manipulation.}
\end{enumerate}
More details can be found in ~\citet{Susobhanan2021} (and references therein). The pipeline parameters used are listed in Table \ref{tab:pipeline}. \bthis{The subsequent analyses used tools provided by \texttt{PSRCHIVE}\footnote{\url{https://psrchive.sourceforge.net/index.shtml}}
software package \citep{Hotan2004,vanStraten2012}. For each
epoch, an optimised  S/N was obtained using the \texttt{pdmp} command
provided by \texttt{PSRCHIVE}.}
The resulting S/N values were used for selecting the optimal observations analysed in this paper.
We selected five epochs from each cycle for both Band 3 and Band 5, except for Band 5 in Cycle 44,  since we did not have enough high S/N epochs.
We also analysed six additional  epochs, which have moderate S/N ratio in Band 3, and which were already selected for Band 5, to check the difference between Band 3 and 5 in the same epoch.
In total, we analysed 26 epochs for Band 3 and 18 epochs for Band 5 as listed in Table \ref{tab:band3_observation} and \ref{tab:band5_observation}, respectively.


We first  collapsed all the frequency channels in each PSRfits file into eight sub-bands after de-dispersing them, using the \texttt{pam} command of \texttt{PSRCHIVE}. \bthis{Here, \texttt{pam} (which stands for Pulsar Archive Manipulator) is one of the commands within \texttt{PSRCHIVE}, which  is  widely used for post-processing partially folded profiles, such
as dedispersion, collapsing the data over frequency and
time and modifying the meta-data in the PSRFITS header.}
\bthis{ Next, the time-of-arrival (ToA) for each partially folded profile was
obtained by cross-correlating these with a frequency resolved
template. The template was formed by averaging all the
sub-intergations using \texttt{pam} while preserving the
eight subbands. These time-collapsed data were then
dedispersed to obtain a final noise-free template
using a wavelet filter implemented in \texttt{psrsmooth}
command of \texttt{PSRCHIVE}. ToAs for every
sub-integration in each observations were then
obtained by  cross-correlating the partially
folded profiles with the noise-free template
using the \texttt{pat} command of \texttt{PSRCHIVE}. }

\bthis{ Finally, we obtained the timing residuals using the
\texttt{TEMPO2}\footnote{\url{https://bitbucket.org/psrsoft/tempo2/src/master/}}
software packages for further analysis described in the
next section.} \bthis{\texttt{TEMPO2} is a package for pulsar
timing analysis. This software compares the observed ToA with that
predicted from a timing model, consisting of rotational,
astrometric and binary parameters of a pulsar ~\citep{Hobbs2006, Hobbs2}.
The sum of squares of the differences between the observed
and predicted ToAs, called timing residuals, is minimised
to obtain the best fit parameters of the pulsar. \texttt{TEMPO2}
is also used for simulating the TOAs using plug-ins provided
in the software. We used \texttt{TEMPO2} for simulations
to estimate pulse jitter as explained in the next section. } 
Before proceeding with the jitter measurements, we removed the TOAs with large uncertainties from a visual check. Finally, we carried out parameter fitting using only the spin frequency and DM, since the observations typically span 10 minutes in duration.

\section{Jitter measurements}
\label{sec:jitter}
In this section, we first describe the noise models and the methods used to measure the jitter amplitudes.
Thereafter, we  present the results of jitter measurements.

\begin{figure}[t]
    \centering
    \includegraphics[width=0.9\linewidth]{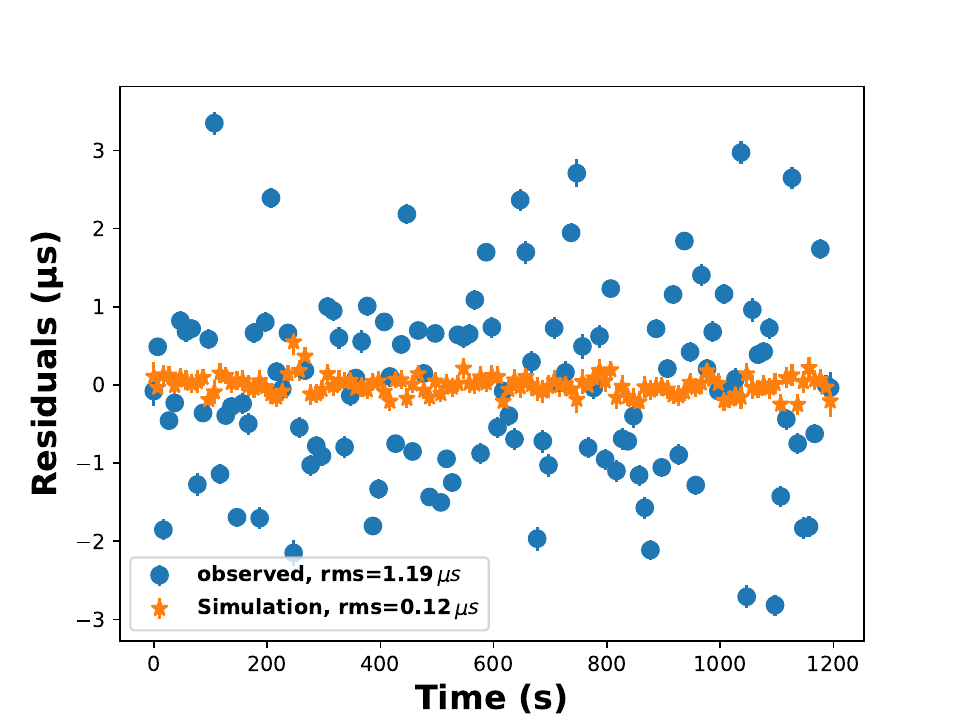}
    \caption{An example of the frequency-averaged timing residuals. The blue circles and orange stars denote the observed residuals obtained from 10 s sub-integrated profiles and residuals obtained with \texttt{libstempo} simulation, respectively (see the beginning of Section \ref{sec:jitter}). The TOAs are obtained from Cycle 41 Band 3 observations for MJD 59545.}
    \label{fig:residuals}
\end{figure}

Traditionally, the jitter amplitudes have been estimated by computing the quadrature difference of the root-mean-square (rms) of frequency averaged residuals, $\sigma_\mathrm{obs}$ and the rms expected from the  radiometer noise, ($\sigma_\mathrm{rad}$):
\begin{equation}
    \sigma_\mathrm{J}^2(T) = \sigma_\mathrm{obs}^2(T) - \sigma_\mathrm{rad}^2(T), 
    \label{eq:sigma_J}
\end{equation}
where $T$ is the length of a sub-integration.
We can estimate $\sigma_\mathrm{rad}$ by considering the Gaussian noise expected from the observed TOA uncertainties.
The simulated TOAs can be obtained using the \texttt{fakepulsar} method from the \texttt{libstempo}\footnote{\url{https://github.com/vallis/libstempo/}} package, with the frequency averaged TOAs and errors as input. \bthis{\texttt{fakepulsar}  is a plug-in within the {\tt TEMPO2} package. This particular plug-in
enables the user to create simulated TOAs that fit a given timing
model in the form of a given parameter file~\citep{Hobbs2006}. The timing
model may correspond to either a real or a hypothetical pulsar. The
addition of red and white noise is possible in this plug-in.}
The simulated timing residuals are encoded in an array with the same length   as  the total number of TOAs, which is filled with zeros. Then we can obtain the timing residuals containing radiometer noise by adding Gaussian noise using \texttt{add\_efac} option. The rms of the timing residuals is denoted by $\sigma_\mathrm{rad}$. We obtain $\sigma_\mathrm{rad}$ with 1000 realisations and then calculate the jitter amplitude $\sigma_\mathrm{J}$ using equation \ref{eq:sigma_J}.
One example of the frequency averaged timing residuals and the simulated residuals is shown in Figure \ref{fig:residuals}. 
We can see that the frequency averaged timing residuals have large fluctuations  compared to the simulated timing residuals,  
suggesting that the remaining TOA fluctuation is due to pulse jitter.  
Jitter noise behaves like a white noise source and hence its amplitude scales inversely with  the square root of the integration time, as shown below.
\begin{equation}
    \label{eq:scaling}
    \sigma_\mathrm{J}(T_1) = \sigma_\mathrm{J}(T_2) \sqrt{T_2/T_1} 
\end{equation}
For the rest of the paper, we report the jitter amplitudes, after rescaling them to one hour for the ease of comparison with previous studies. 

\subsection{Noise model}
\label{sec:noisemodel}
For the purpose of this study, we use Bayesian inference to estimate the jitter amplitude and used the traditional method described in the beginning of this section for crosscheck. 
In this subsection, we briefly describe the noise model used to estimate the jitter amplitude in this work.
The following description is based on previous works ~\citep{Lentati2014,HV14,Arzoumanian2015,Srivastava23}, where more details can be found. 

The observed timing residuals can be defined in terms of an $N_\mathrm{TOA}$ length vector, where $N_\mathrm{TOA} = N_{\mathrm{subint}} \times N_\mathrm{subband}$ 
  for a given observation, and can be modelled as a sum of various deterministic and stochastic noise sources. Since we are interested in the stochastic noise within a single epoch, we assume that the timing residual vector ($\delta$t) can be fully described by the white noise term after removing the deterministic term.
Therefore, the likelihood function can be  written as follows:
\begin{equation}
    \label{eq:likelihood}
    L(\delta \mathbf{t} | \theta) = \frac{1}{\sqrt{(2\pi)^{N_\mathrm{TOA}} \det(\mathbf{C}(\theta)) }}\exp\left(-\frac{1}{2}\delta\mathbf{t}^T \mathbf{C}(\theta)^{-1} \delta\mathbf{t} \right), 
\end{equation}
where $\mathbf{C}(\theta)$ is the covariance matrix and $\theta$ is a set of parameters. 


In pulsar timing analysis, there are three parameters that are used to characterize white noise: EFAC, EQUAD, and ECORR.
EFAC is a multiplicative scale  factor that corrects for the underestimation  of  the TOA uncertainty that may be due to errors in calibration.
EQUAD is a term that adds white noise in quadrature to represent the range of fluctuations beyond the TOA uncertainty.
ECORR is similar to EQUAD, where  white noise is again added in quadrature, but ECORR is uncorrelated in  different time bins, but perfectly correlated at different frequencies.
It represents the fluctuation of the  TOA due to pulse profile variation.
Therefore, ECORR corresponds to the jitter noise.
EQUAD  also corresponds to jitter as far as a single sub-band TOA is concerned.
In this work, we use EFAC and EQUAD, when we estimate the  jitter amplitude for each sub-banded TOA, and use EFAC and ECORR from the whole set of  TOAs.
We used the following three models to estimate the  jitter amplitude:
\begin{description}
    \item[Model I] common EFAC and ECORR throughout the band.
    \item[Model II] common EFAC throughout the band and different EQUAD for each sub-band.
    \item[Model III] different EQUAD for each sub-band without EFAC.
\end{description}
In Model I, ECORR corresponds to the jitter amplitude for the given band.
In  Model II and III, we estimate the jitter for each sub-band.  The aim of Model III is to eliminate any correlation between the sub-bands, since we used a  common EFAC in Model II.
The covariance matrix for each model can be  written as follows:
\footnote{
In the modern definition of EQUAD, EQUAD is defined with EFAC multiplied, but since we are interested in measuring the jitter amplitude in Eq \ref{eq:sigma_J} within a  Bayesian framework, we adopt the definition of {\tt temponest} without EFAC applied.
}
\begin{align}
    \label{eq:covariance}
    &\mathrm{Model\, I} \quad 
                \mathbf{C}_{ij} = 
                    F ^2\sigma_{\mathrm{TOA},i}^2\delta_{ij}
                + J^2 \delta_{t_i,t_j} \\
    &\mathrm{Model\, II} \quad
                \mathbf{C}_{ij} = 
                \left(
                    F^2 \sigma_{\mathrm{TOA},i}^2
                    + Q_{\nu_i}^2
                \right)
                \delta_{ij} \\
    &\mathrm{Model\,III} \quad 
                \mathbf{C}_{ij} = 
                \left(
                    \sigma_{\mathrm{TOA},i}^2 + Q_{\nu_i}^2
                \right)
                \delta_{ij}
\end{align}
where 
$\sigma_{{\mathrm{TOA}},i}$ is the uncertainty of the $i$th TOA.
$F,\, J$ refer to  EFAC and ECORR, respectively.
$Q_{\nu_i}$ denotes EQUAD for the $i$th sub-band.
The index $t_i$ represents the sub-integration of the $i$th TOA to incorporate ECORR in the same sub-integration.
We used uniform prior distribution in [0.5, 5.0] for  EFAC  and log-uniform prior distribution in [-10.0, -5.0] for  EQUAD and ECORR.
\begin{figure}[t]
    \centering
    \includegraphics[width=0.9\linewidth]{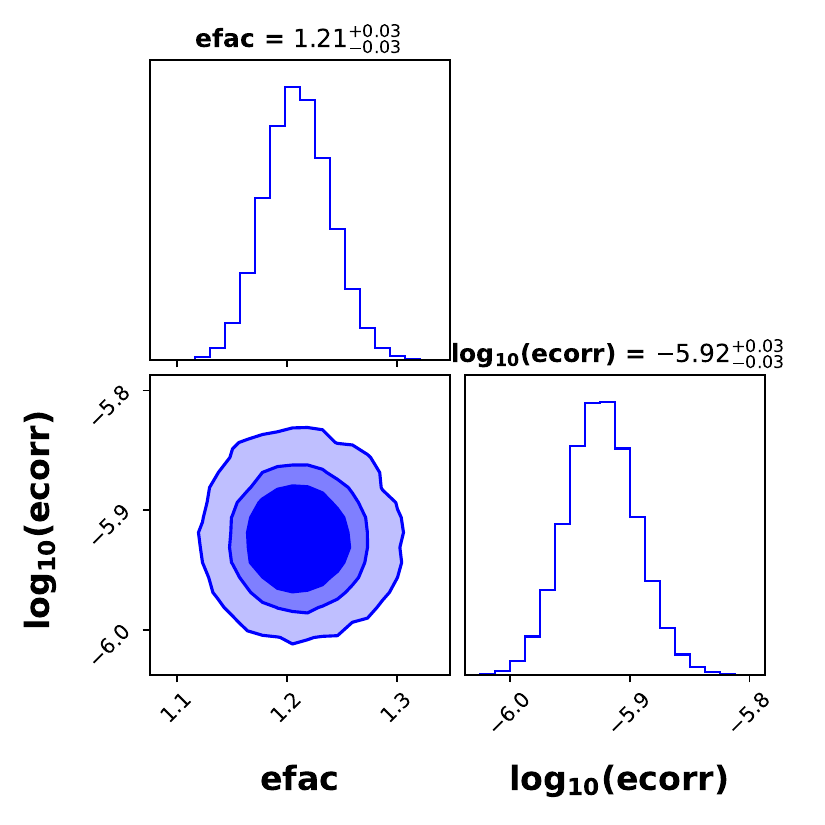}
    \caption{An example of the posterior distribution with 68\%, 90\%, 99\% credible intervals  of the Model I parameter estimation. The point estimates  shown above the plots represent the median values and marginalized  68\% credible intervals. The TOAs are obtained from Cycle 41 Band 3 observations on  MJD 59545.}
    \label{fig:efac_ecorr}
\end{figure}

\subsection{Data Analysis}
\label{Sec:DataAnalysis}
To measure the  jitter amplitude using Bayesian inference, we used the \texttt{libstempo} and the \texttt{ENTERPRISE} python packages \citep{enterprise}. \bthis{\texttt{libstempo} is a Python wrapper to the \texttt{tempo2} package.} 
\bthis{\texttt{ENTERPRISE} is a pulsar-timing analysis software for pulsar noise analysis, GW searches and pulsar timing model analysis.}
We used the sub-banded TOAs and the corresponding {\tt par} file as input. We then define the likelihood function depending on the Model I-III, from which the noise parameters are then estimated.
We used the  Markov Chain Monte-Carlo (MCMC)  sampler implemented in the \texttt{PTMCMCSampler} package \citep{ptmcmc} to sample  the  posterior  and estimate the EFAC and ECORR or EQUAD. \bthis{\texttt{PTMCMCSampler} is an acronym for Parallel Tempering Markov Chain Monte-Carlo  sampler, and  utilizes an adaptive jump proposal mechanism by default, incorporating both standard and single-component Adaptive Metropolis (AM) as well as Differential Evolution (DE) jumps. Moreover, \texttt{MPI (mpi4py)} is employed to execute the parallel chains in this implementation.}
One  example of such a  posterior plot along with the marginalized posteriors for EFAC and ECORR  is shown in Figure \ref{fig:efac_ecorr}.

\begin{figure}[t]
    \centering
    \includegraphics[width=0.9\linewidth]{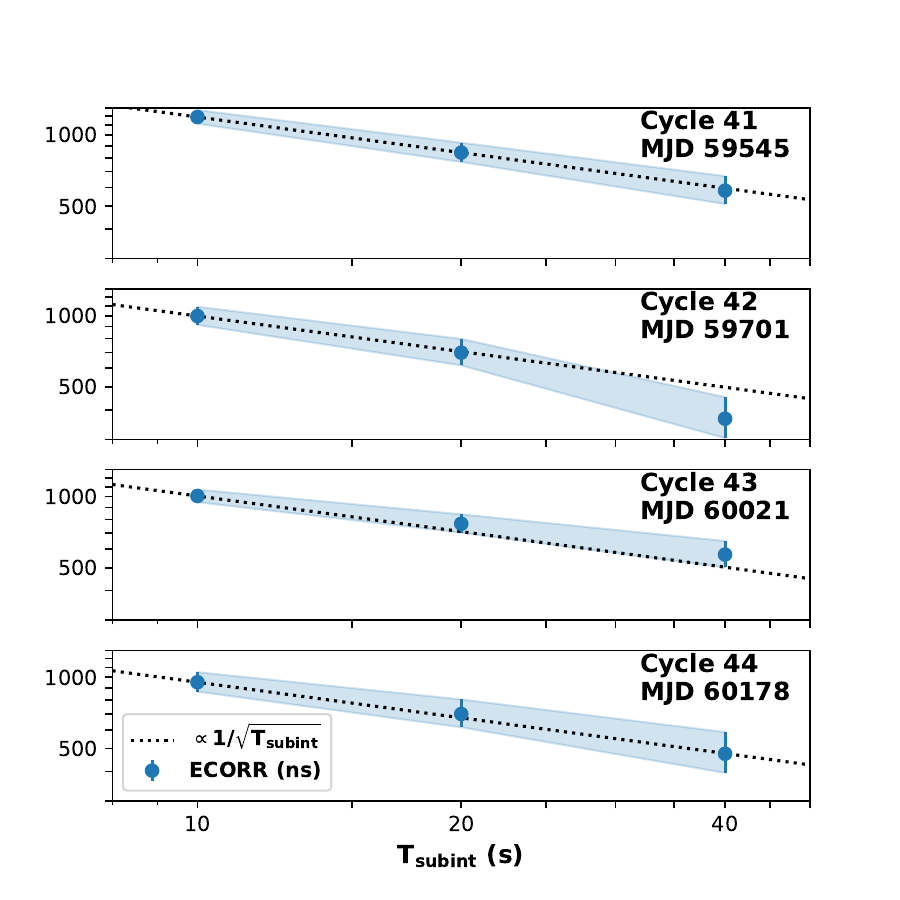}
    \caption{ECORR obtained with different sub-integration.  Blue circles denote the  ECORR values obtained from different sub-integration times. The black dotted lines shows the fit, which is   proportional to the inverse square root of the sub-integration times. }
    \label{fig:scaling_law}
\end{figure}

As mentioned at the beginning of Section \ref{sec:jitter}, we rescale the estimated  values of EQUAD and ECORR to one hour duration, using the same scaling relation as in equation \ref{eq:scaling}. However, in order to reaffirm whether they would  follow the same relation as equation \ref{eq:scaling}, we recalculate ECORR for Model I by changing the sub-integration time to 20 and 40 seconds.
The result is shown in Figure~\ref{fig:scaling_law}, and we confirm that  ECORR also follows the correct scaling law in accord with equation~\ref{eq:scaling}.
Although we do not have data for observation durations longer than one hour, we assume that the scaling law would hold even if the integration time is extended to one hour.
Therefore, we scaled the jitter amplitude obtained by Bayesian analysis to one hour for ease of comparison with previous studies. 

\begin{figure}[t]
    \centering
    \includegraphics[width=0.95\linewidth]{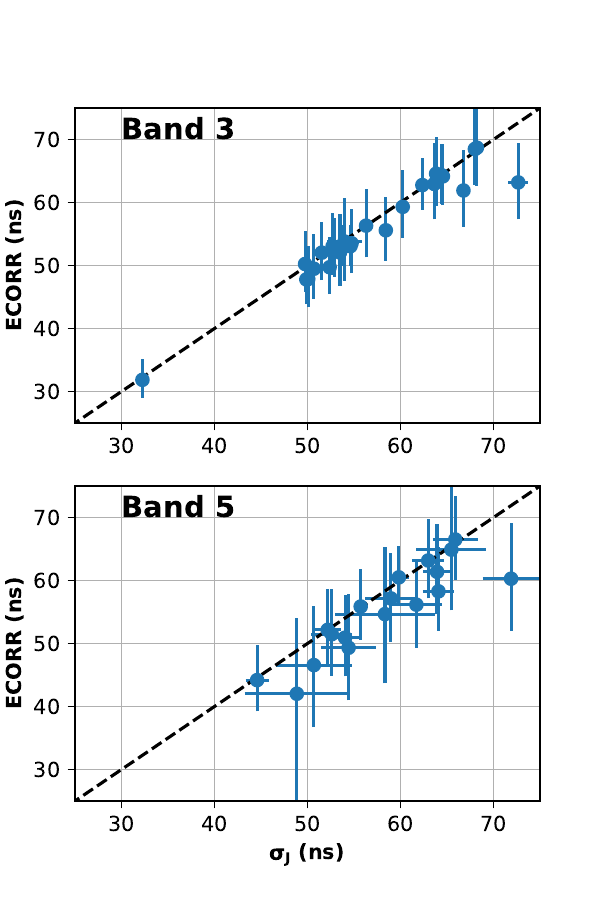}
    \caption{\bthis{Comparison of jitter amplitudes obtained from Bayesian analysis and traditional method. The blue points denote jitter amplitude. The dashed black line represents the straight line given by $y=x$. All values are scaled to 1 hour value using equation \ref{eq:scaling}.}}
    \label{fig:consistency_check}
\end{figure}

As a consistency check, we also 
verify whether the ECORR values obtained with {\tt ENTERPRISE} 
are consistent with the jitter amplitude obtained with the traditional method described in the beginning of this section
\bthis{(see Figure \ref{fig:consistency_check}). }
We find that all the $\sigma_\mathrm{J }$ values
agree with the ECORR values to within $1\sigma$, thereby showing that our results are self-consistent.

%

\subsection{Jitter amplitude within the band}
\label{sec:jitteramplitude}
\begin{figure*}
    \centering
    \includegraphics[width=0.95\linewidth]{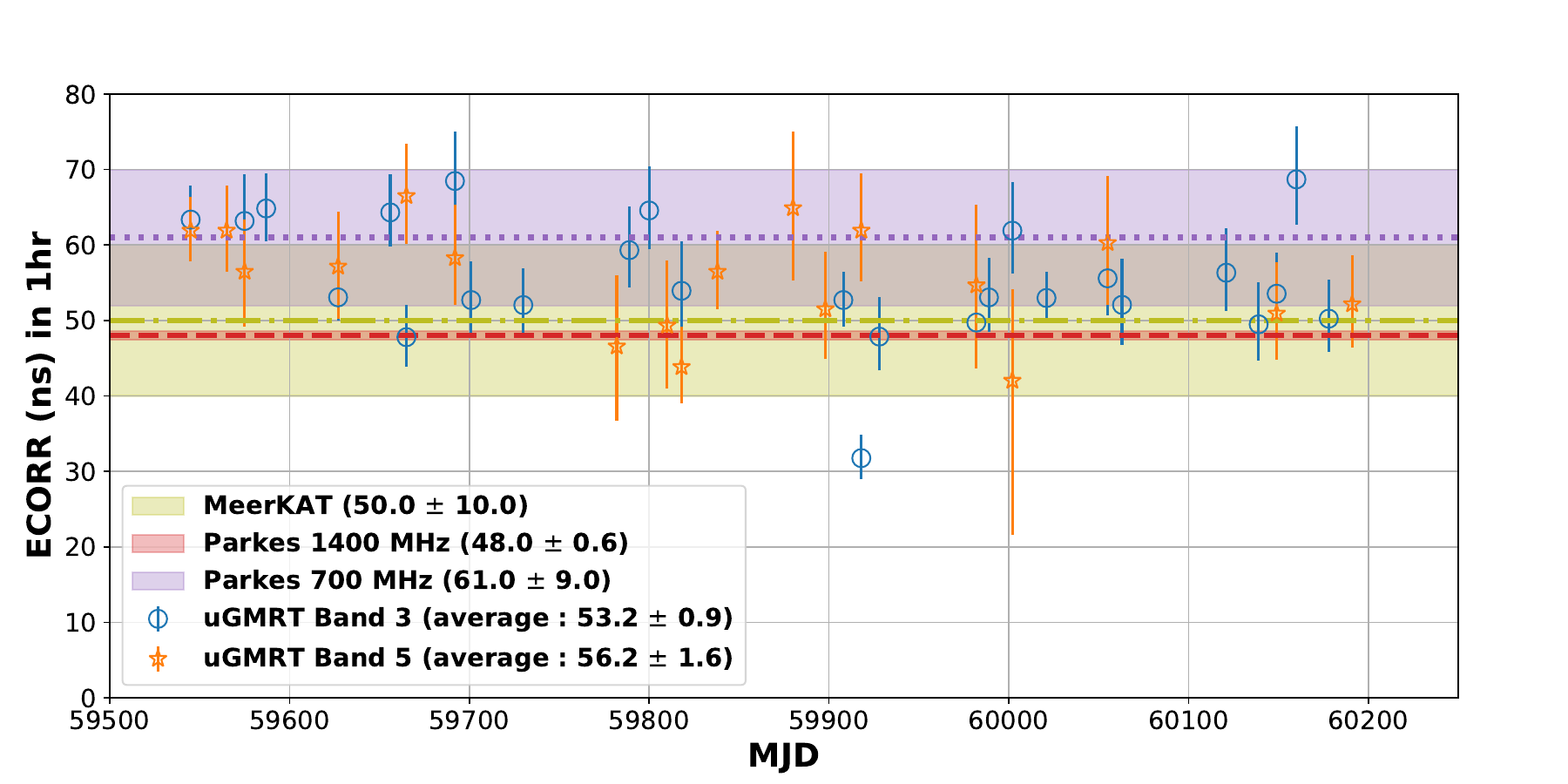}
    \caption{ECORR time series scaled to one hour obtained from Model I estimation. The blue circles and orange stars denote the  ECORR values obtained from Band 3 and Band 5 observations, respectively. The purple shaded region  and the dotted line show the  results of MeerKAT observations~\citepalias{Parthasarathy2021}. The red shaded region and the dashed line and the yellow shaded region and dot-dashed line show the corresponding results for the Parkes observations~\citepalias{Shannon2014}. Note that the values shown in the previous studies are not ECORR, but $\sigma_\mathrm{J}$ defined by the equation \ref{eq:sigma_J}, which is measured by the traditional method described in the beginning of Section \ref{sec:jitter}.}
    \label{fig:ECORR_timeseries}
\end{figure*}

\begin{figure*}
    \centering
    \includegraphics[width=0.95\linewidth]{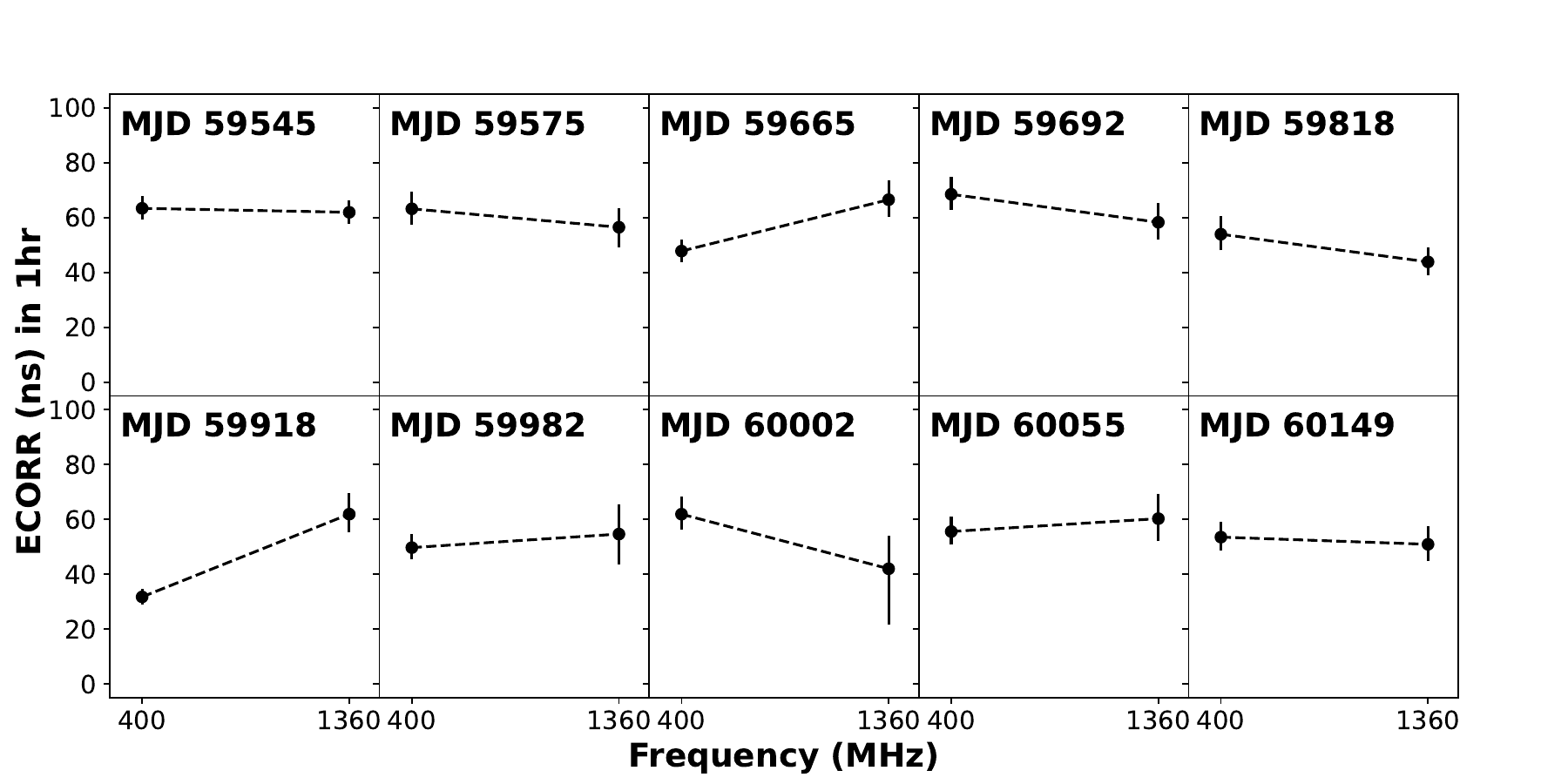}
    \caption{Comparison plots of ECORR for Band 3 and Band 5 for the same epoch obtained from Model I estimation.}
    \label{fig:BAND35_comparison}
\end{figure*}
Using the TOAs obtained by the method described in Section~\ref{sec:DataProcessing}, we obtained ECORR values for all the selected epochs which are listed in Tables \ref{tab:band3_observation} and \ref{tab:band5_observation}, for Model I of Section \ref{sec:noisemodel}, which considers ECORR as a proxy for  the jitter amplitude throughout the band. The resulting values of ECORR  are rescaled to one hour and are listed in column 5 of Tables \ref{tab:band3_observation} and \ref{tab:band5_observation}.

The rescaled ECORR values are plotted in Figure \ref{fig:ECORR_timeseries} for each epoch. We estimate the weighted average of ECORR to be  $53.2 \pm 0.9$ ns in Band 3 and $56.2 \pm 1.6$ ns in Band 5 from our observations.
We find that our Band 5 result is consistent with \citetalias{Parthasarathy2021}. The jitter amplitude in \citetalias{Shannon2014} was reported for  a single observation only. Though our weighted average value for Band 5 differs from that quoted in \citetalias{Shannon2014}, the spread in our overall results  ensures that we are in good agreement with \citetalias{Shannon2014} as well.


Our results for ECORR at Band 3 are the first jitter measurements calculated for PSR~J0437$-$4715 below 700 MHz. Therefore, we do not  have previous studies to directly compare our results. The frequency dependence suggested by \citetalias{Shannon2014} and \citetalias{Parthasarathy2021} point to a higher jitter at lower frequencies. The results we derive are, however, not in agreement with this conclusion since we find that  ECORR at Band 3 is consistent with that seen in Band 5. Also, we see that the weighted average of the jitter amplitude at Band 3 is slightly lower than that of Band 5. We calculated the weighted average again after excluding the epoch corresponding to MJD 59918, since it is likely to be an outlier as it is strongly affected by scintillation. This gave a new weighted average value of jitter at Band 3, equal to $55.47 \pm 0.94$ ns. Therefore, even  after removing the  outlier, 
the jitter in Band 3 is consistent with that in  Band 5 within $1\sigma$. 
One possible reason that the jitter in  Band 3 is  not so large could be due to  epoch to epoch variations, since we are comparing different epochs. In Figure \ref{fig:BAND35_comparison}, we plot the ECORR values for Band 3 and Band 5 from same epochs, for comparing their behaviour. We see that while there are some epochs where the jitter in Band 3 is larger, there are also epochs where the jitter values at Band 3 are the same or smaller. From our whole band jitter (ECORR) estimates, we conclude that it is not possible to say with certainty whether the jitter is larger or smaller at lower frequencies than at higher frequencies.


We see that there is a large scatter in the Band 3 jitter estimates. On investigating the Band 3 pulsar profiles in the  frequency domain, we find that  this scatter in the jitter values might be caused by the scintillation present in many profiles. Further, we notice that in this band, the S/N varies significantly with frequency as well as from epoch to epoch, which  again can possibly be explained by the presence of interstellar scintillation in the data. This may have caused the failure in measuring the fluctuations of the timing residuals for some of the sub-bands which have low-S/N, thereby resulting in higher jitter estimates. As a result, we further scrutinize the  Band 3  data for further insights on the nature of jitter.

\subsection{Intra-Band Frequency dependence of Jitter}
\label{sec:frequencydependence}
\begin{figure}[t]
    \centering
    \includegraphics[width=0.9\linewidth]{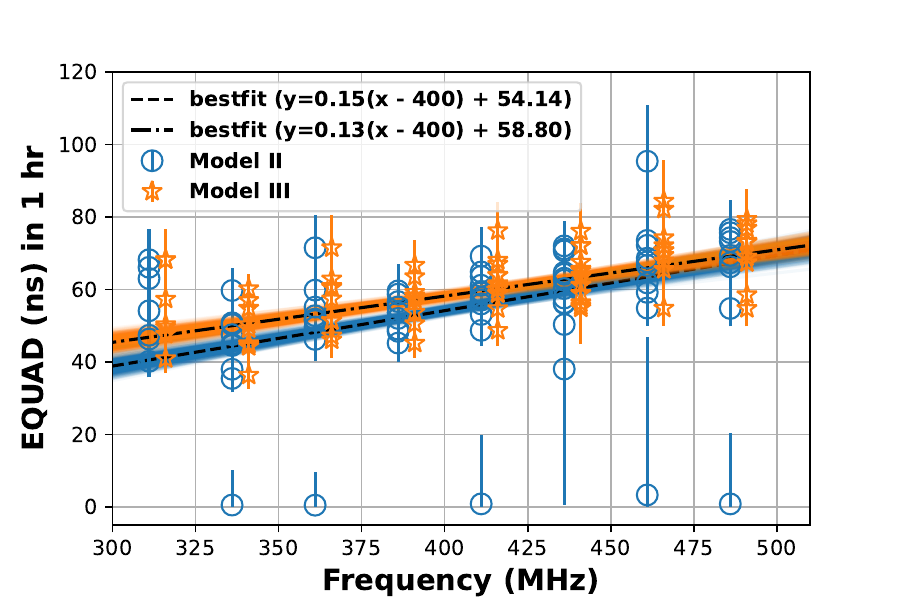}
    \caption{Sub-banded EQUAD scaled to one hour obtained from Model II and III estimation for the sub-bands which had the $\mathrm{S/N}_\mathrm{sub}$ above the threshold value. The blue circles and orange stars represent the  sub-banded EQUAD value of Model II (with EFAC) and III (without EFAC), respectively. The orange stars have been  offset by 5 MHz to the right for clarity. The blue and orange shaded lines are from 500 random lines obtained from the posterior distribution of Bayesian regression. The reduced $\chi^2$ of the best-fit lines are equal to  4.8 for Model II and 1.7 for Model III.    
}
    \label{fig:equad_subband}
\end{figure}

\begin{figure}[t]
    \centering
    \includegraphics[width=0.9\linewidth]{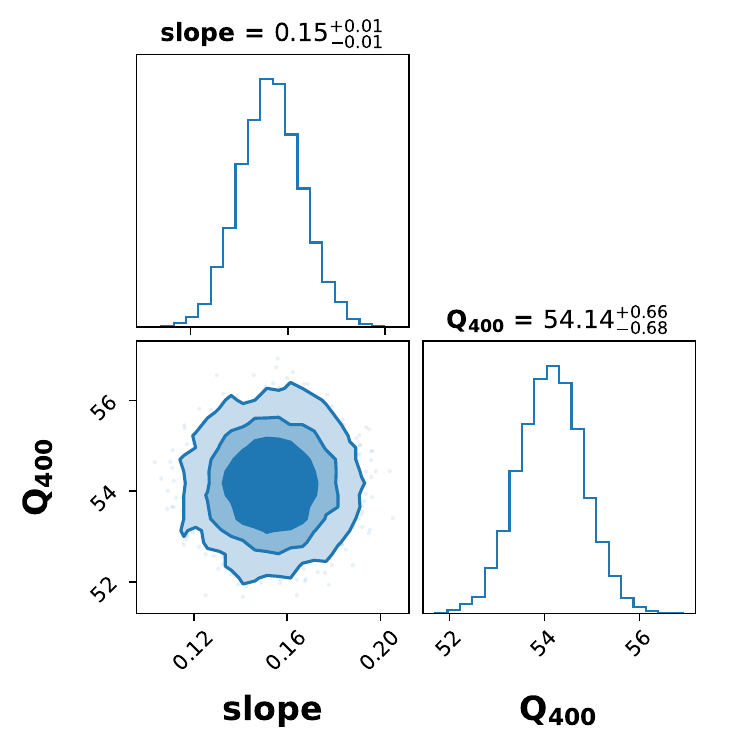}
    \caption{Posterior distributions with 68\%, 90\%, 99\% credible intervals for the  parameters fitted using Model II in Figure \ref{fig:equad_subband}.  The point estimates  shown above the plots represent the median values and marginalized  68 \% credible intervals. }
    \label{fig:withEFACregression_posterior}
\end{figure}

\begin{figure}
    \centering
    \includegraphics[width=8cm]{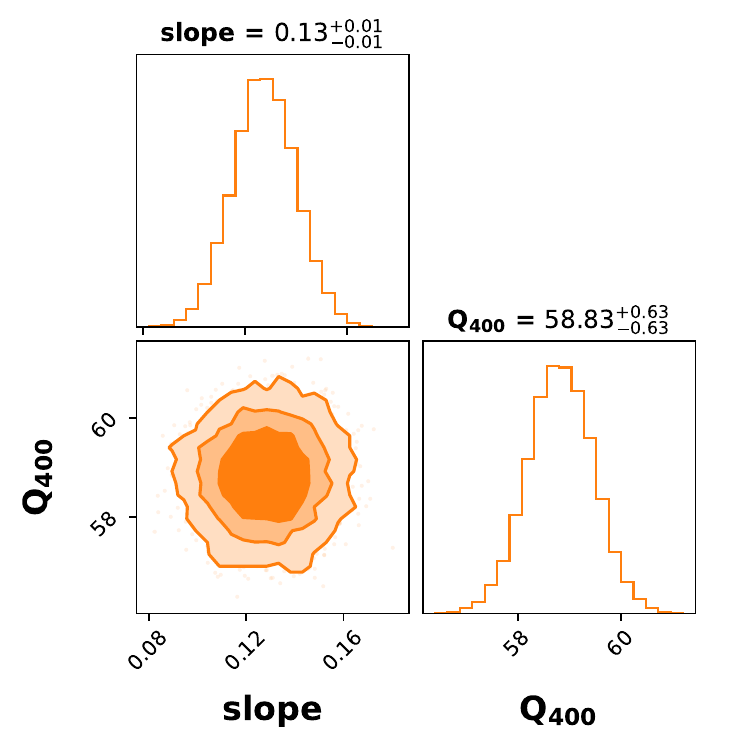}
    \caption{Posterior distributions with 68\%, 90\%, 99\% credible intervals  for the  parameters fitted using Model III in Figure \ref{fig:equad_subband}. The point estimates  shown above the plots represent the median values and marginalized  68 \% credible intervals. }
    \label{fig:withoutEFACregression_posterior}
\end{figure}


In Section \ref{sec:jitteramplitude}, we have demonstrated  
that the jitter estimates at low frequency may be affected by scintillation. This prompted us to look at the sub-banded data to get additional insight into the nature of jitter in the different frequency channels of Band 3. In this analysis, we estimate EQUAD for each sub-band using Model II and III from the 20 highest S/N epochs listed in Table \ref{tab:band3_observation}. EQUAD is used to account for  the unaccounted (besides telescope noise) noise in the TOA uncertainty, and it corresponds to the  jitter amplitude estimated using the traditional method described in the beginning of Section \ref{sec:jitter} \citepalias{Shannon2014,Parthasarathy2021}.

Learning from our experience in Section \ref{sec:jitteramplitude}, where low S/N hindered us from getting  accurate jitter estimates at Band 3, we obtain credible EQUAD values by first calculating the  sub-banded S/N and then adopting EQUAD values  from only those high S/N observations that lie above a given threshold.
The sub-banded S/N is defined as follows:
\begin{equation}
    \mathrm{S/N_{sub}} = \frac{\sum_{i, j} I_{i,j}}{\sigma_I}, 
    \label{eq:subbanded_SN}
\end{equation}
where $I_{i,j}$ is the intensity of $i^{th}$ phase bin and $j^{th}$ sub-integration in a certain subband,
$\sigma_I=\sqrt{\sum_{i} \sigma_{I, i}^2}$ is the rms of the intensity in a certain subband, and $\sigma_{I,i}$ is the rms of the intensity in the $i^{th}$ phase bin.
We used the median value of the $\mathrm{S/N_{sub}}$ as our threshold, which is equal to  $\mathrm{S/N_{sub} = 2606}$.


The results from parameter estimation for Model II and III are shown in Figure~\ref{fig:equad_subband}, which clearly show a tendency for an  increase in EQUAD with  frequency for both the models. To reaffirm this quantitatively, we performed a Bayesian regression and estimated the slope ($a$), and EQUAD at 400 MHz ($Q_{400}$)  using the  \texttt{emcee}\footnote{\url{https://github.com/dfm/emcee}} MCMC sampler~\citep{Foreman2013}. The \bthis{emcee  package uses  an  affine invariant Affine Invariant Markov chain Monte Carlo Ensemble sampler using the algorithm in \citet{Goodman}.}
We used the following regression model:
\begin{equation}
    y = a (f-400) + Q_{400}
    \label{eq:lenearfunc}
\end{equation}
where $f$ is the frequency. 
After performing the regression, we randomly sampled 500 values from the posterior distribution of $a$ and $Q_{400}$, 
and then the results using these values are depicted as the blue and orange regions in Figure~\ref{fig:equad_subband}.
We see that the lines have  a positive slope implying an increase in EQUAD with frequency, a trend that is opposite to that of \citetalias{Shannon2014} and \citetalias{Parthasarathy2021}.

The posterior distributions of $a$ and $Q_{400}$ are shown in  Figure~\ref{fig:withEFACregression_posterior} and \ref{fig:withoutEFACregression_posterior}, respectively. 
These plots show that the slopes have a positive value with significance of about $15\sigma$. 
The plots also list the value of $Q_{400}$, or the jitter amplitude at 400 MHz, which is about $54.14^{+0.66}_{-0.68}$ ns, in  Model II and $58.83^{+0.63}_{-0.63}$  ns, in Model III.




\subsection{Jitter and the DM variations}
\label{sec:dm}

\citetalias{Parthasarathy2021} raised the concern of the jitter being the limiting factor in the DM precision measurements from shorter duration observations in all MSPs, and suggested that it can only be overcome by longer integrations, based on their analysis in the  high frequency regime. We investigated the effect of jitter on the DM precision in the  low frequency regime (300 MHz - 500 MHz) by studying the timing residuals and obtaining the DM uncertainties for 10 second sub-integrations from a 20  minute observation done at the uGMRT.

\begin{figure}[t]
\centering
\includegraphics[width=0.9\linewidth]{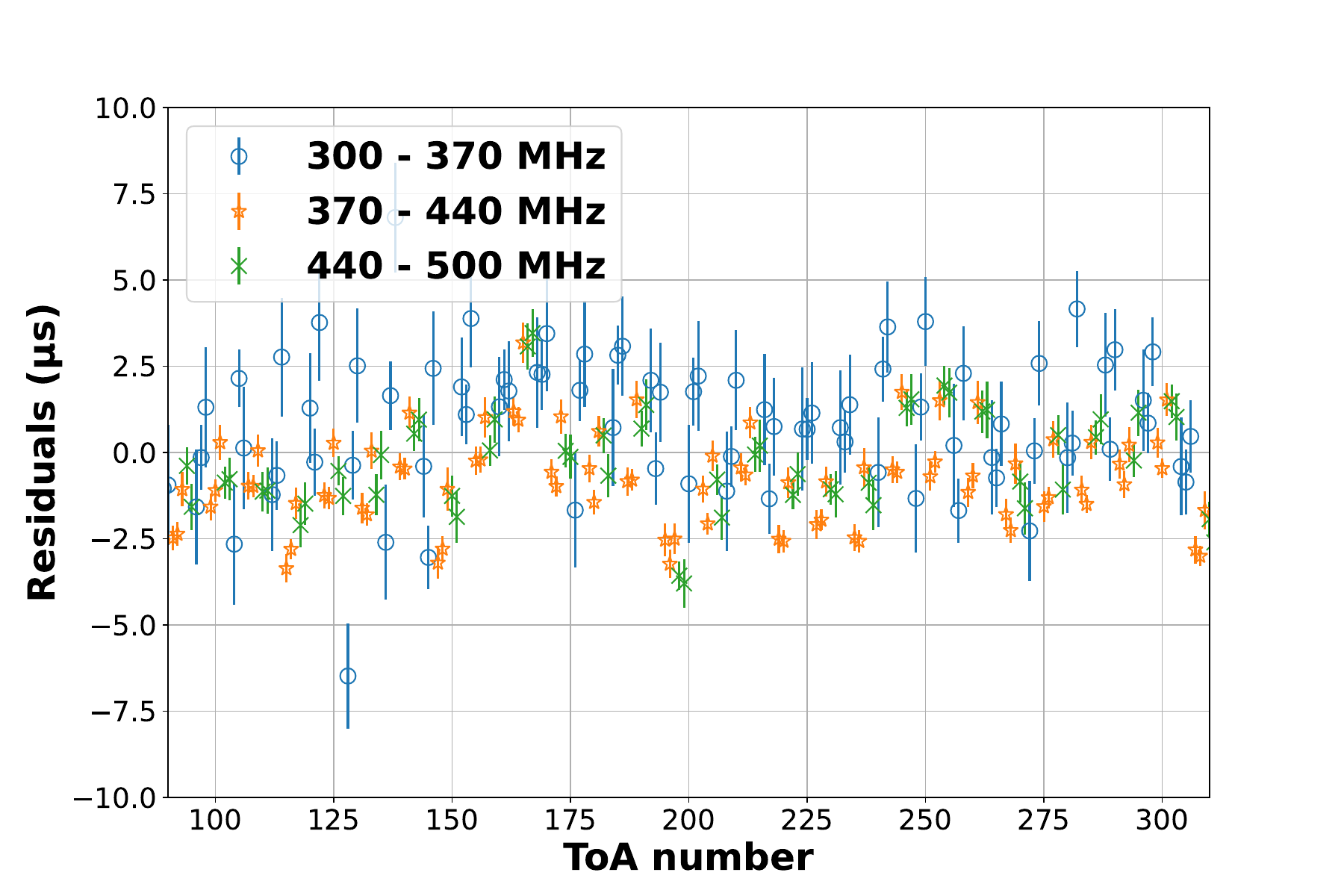}
\caption{A subset of the `Post-fit timing residuals' of PSR~J0437$-$4715 estimated using 10 second sub-integrated profiles with each having eight frequency channels. The TOAs are subdivided and colour coded as three sets based on the frequency sub-band as shown in the figure. These residuals are plotted serially against the TOA numbers, to check for any potential frequency dependence and its variation for each TOA set in the plotted subset. The complete observation spanned 20 minutes,  and only a subset is plotted in this figure for clarity.}
\label{fig:timingresiduals}
\end{figure}

We measured the timing residuals from each of its 10 second sub-integrations, where  each sub-integration is divided into eight  frequency channels. We plot in Figure~\ref{fig:timingresiduals}, a small subset of the  post-fit timing residuals derived from the TOAs. These are plotted serially in time, with each TOA observation being 10 seconds long. The residuals are colour coded for a given frequency sub-band (as mentioned in Figure \ref{fig:timingresiduals}). There seems to be a dependence of the \bthis{timing residuals} on frequency on a 10 second timescale, which may be due to the frequency dependence of DM.  The dependence of the TOA on frequency at Band 3 is, however, not as strong as that  shown for the  high frequency regime  in \citetalias{Parthasarathy2021}. This probably implies a smaller spread in DM at Band 3 compared to that in the higher frequency band reported in \citetalias{Parthasarathy2021}.

\begin{figure}[t]
\centering
\includegraphics[width=0.9\linewidth]{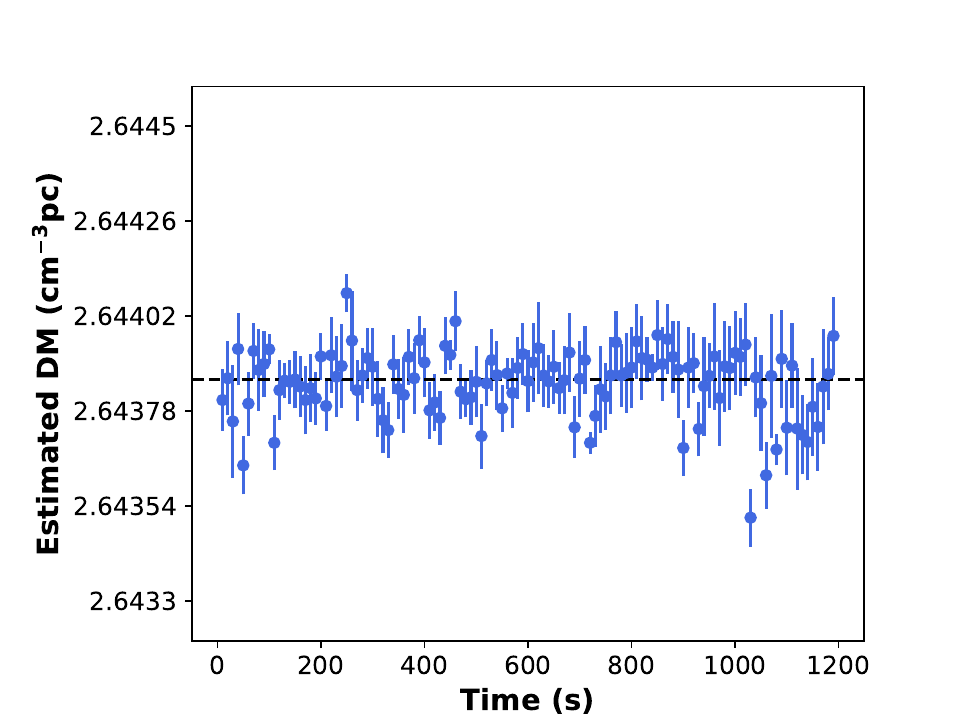}
\caption{The estimated DM values for each 10 second sub-integration from {\tt psrfits} to the timing residuals shown in Figure \ref{fig:timingresiduals}. The horizontal dotted line represents the median estimated DM value of 2.64386 $\rm{cm^{-3}   \,pc}$ 
}
\label{fig:dm}
\end{figure}

To explore this, we measured the  DM values independently from the TOAs of each 10 second sub-integration of the observation. From this,  we estimated the   median DM to be  $2.64386$ $\rm{cm^{-3}   \,pc}$ and  a standard deviation  of $8.4 \times 10^{-5} \rm{cm^{-3}   \,pc}$. This is shown graphically in Figure \ref{fig:dm}. We conclude that the scatter in DM is less at Band 3 compared to higher frequency bands (see figure 4 of \citetalias{Parthasarathy2021}).


Further, to understand the effect of jitter on the DM precision measurements and to get one more independent check on the DM estimation from observations alone, we use the following equation to derive the resultant error in the  DM estimation:
\begin{equation}
    \label{eq:dm_eq}
    {\Delta t}= 4.15 \times 10^6 (\rm{ms}) \left(\frac{1}{ \it{{f_1}{^ 2}} } - \frac{1} { \it{{f_2}{^ 2}} }\right)DM,
\end{equation}
where $\Delta t$ is the delay in TOAs at the two observing frequencies, $f_1$ and $f_2$. 

As shown in Sections \ref{sec:jitteramplitude} and \ref{sec:frequencydependence}, the pulse jitter is frequency dependent. In Band 3, it has a larger value at  higher frequencies, as can be seen from figure \ref{fig:equad_subband}. To estimate  the contribution of pulse jitter in the DM precision, we calculate jitter for a 10 second sub-integration at the  two extreme frequencies of our observation, namely 311 MHz and 486 MHz, using equation \ref{eq:scaling}. We find  the jitter estimate for a 10 second observation to be 1121.69 ns and 1287.87 ns for 311 and 486 MHz, respectively. Assuming the error in TOAs to be due to jitter alone, we infer the DM error to be $\sim 6.7\times10^{-5} \mathrm{cm^{-3} \,pc}$ which is  marginally less than our observationally derived  error of $\sim 8.4\times10^{-5}\mathrm{cm^{-3} \,pc}$. However, the error in the  TOAs is not due to jitter alone and the other major factor which contributes to this error is the telescope noise. This observation on which the analysis is done is a high S/N observation, with a S/N of at least 1000 for each 10 second sub-integration. Hence in this case, the telescope noise is about $\sim 10^{-6}$, which is slightly less than  the jitter noise of  $\sim  1.71\times10^{-6}$,  and therefore, we can assert that the error in TOAs is dominated by the  jitter noise in high S/N \textit{but} lower integration time observations. Including the contribution of telescope noise in the aforementioned analysis increases the DM error marginally to be  around $7.8\times10^{-5} \mathrm{cm^{-3} \,pc} $. Thus, our two independent analyses for the DM uncertainties are in good agreement  with each other, thereby providing confidence in our inferences for the jitter measurement and its effect on the DM precision.

\section{Discussion}
\label{sec:discussion}
\subsection{Origin of the frequency dependence in jitter}
\begin{figure}[t]
    \centering
    \includegraphics[width=0.9\linewidth]{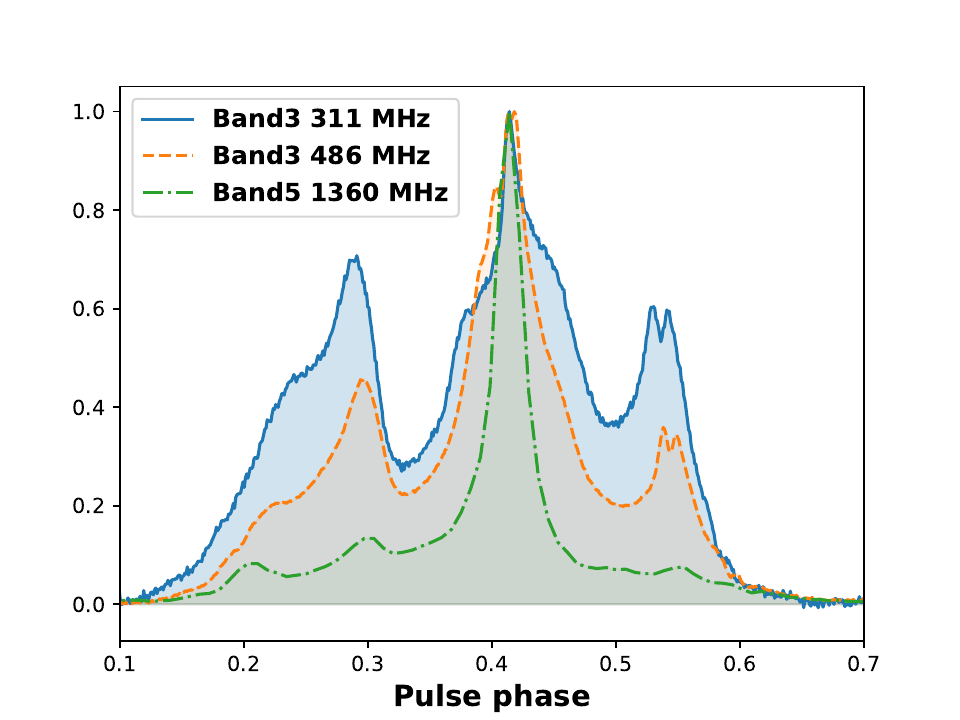}
    \caption{The integrated pulse profile of PSR~J0437$-$4715 at different frequencies. Intensities are normalized. The data used for this plot were obtained from Cycle 41 observation on MJD 59545.}
    \label{fig:profile}
\end{figure}

Our analysis of the high S/N Band 3 observations of PSR~J0437\\$-$4715 reveals a positive correlation of jitter with frequency, as shown in Section \ref{sec:frequencydependence}. 
Therefore, we find an  opposite trend than that suggested by \citetalias{Shannon2014} and \citetalias{Parthasarathy2021} for their higher frequency observations. It could  be due to two possible reasons. One is due to the possibility that the pulse profile is more stable at lower frequencies than at higher frequencies in Band 3. This is because jitter is caused by the pulse profile variation in the pulse-to-pulse scale. And secondly, there seems to exist a \textit{turnover} frequency region at which the jitter trend reverses. This region seems to exist between 500 MHz and 900 MHz, probably around 700 MHz.

The existence of a turnover frequency region at which jitter behavior changes, can possibly have two interpretations, which we outline below.
The first interpretation comes from the self-absorption effect in the pulsar magnetosphere. It has been shown that the high frequency radio emission from  the pulsar comes from  regions closer to its surface, while the low frequency emission comes from regions farther from the surface \citep{Komesaroff1970,Cordes1978}.
This is called the radius-to-frequency mapping (RFM). Of the many pulsars studied, most show this trend but there are a few some exceptions~\citep{Posselt2021}.
Based on the RFM studies, we can say that since the distance of propagation through the  emission region is shorter for lower frequencies, these are less sensitive to  the self-absorption effects, thereby resulting in more stable profiles than those at higher frequencies. If this is true, the \textit{turnover} frequency of the jitter amplitude could be related to the spectral \textit{turnover} of the pulsar flux density.
However, \citet{Lee2022} reported two \textit{turnover} frequencies of the flux density for PSR~J0437$-$4715, which are $\sim 285 \,\mathrm{MHz}$ and $\sim 1900 \,\mathrm{MHz}$. This is therefore, not supporting our  {\it ansatz} for the change in the jitter trend. In any case, it remains to be seen how jitter behaves around these two frequencies. A holistic mapping of the  jitter trend  over the full span of  frequency bands will probably provide a clearer picture.

The second possibility comes from the profile evolution studies. The analysis in \citetalias{Shannon2014} and \citetalias{Parthasarathy2021} suggested that the possible reason for the smaller jitter at higher frequencies is due to narrower pulses and the larger modulation of the main pulse component at those frequencies. However, our jitter trend is opposite to their results. PSR~J0437$-$4715 seems to have a narrow central component in its profile above 700 MHz \citep{Dai2015}. Comparing 1400 MHz observations with 327 MHz observations for this pulsar, we find relatively more stable precursor and postcursor components at lower frequency~\citep{Osłowski2014,Vivekanand1998}. We notice that the pulse profiles show more significant precursor and postcursor components in our Band 3 observations as well (see Figure~\ref{fig:profile}).  This leads us to posit that the reason for the smaller jitter at lower frequencies at Band 3 is due to the fact that  these precursor and postcursor components are more stable, and hence they moderate the fluctuations in the TOAs by the central (main) component. To investigate this further we need to study the phase-resolved modulation index of single pulses, which may help resolve this issue.
Hence, it will be worth revisiting single pulse studies of PSR~J0437$-$4715 with our high-sensitivity, wideband observations, which will be defered to a future paper.

\subsection{Effect of jitter on the DM precision measurements}

From our analysis, in Section \ref{sec:dm}, we see that the DM precision is limited by jitter noise for high S/N observations on short timescales. Further, if we extrapolate the results from \citetalias{Parthasarathy2021}, it would indicate that we cannot get high precision DM just by going to lower frequencies. However, our analysis suggests otherwise, and it might be possible to get even better precision DMs by going to lower frequencies ($\sim$100 MHz). Though the possibility of low frequency giving better DM precision is supported by our analysis at Band 3, it remains to be seen observationally what trend is followed by pulse jitter at that frequency regime.

Further, our analysis in tandem with  \citetalias{Parthasarathy2021} suggests that it is not possible to get DM precision measurements using observations at \textit{two} (high and low) frequencies only. This is because the DM will depend on the jitter coming from both the observed frequencies; and it is entirely possible that those two frequencies may show entirely opposite trends in jitter (e.g., at Band 3 and Band 5). For more precise estimates of jitter, we need to be able to extrapolate over multiple bands in a frequency resolved manner, since different frequency bands are likely to give different estimate for order of precision.

A comprehensive understanding of the pulse jitter behaviour over different frequency bands and its effect on DM is warranted for better precision pulsar timing. It also remains to be seen what trend in jitter is followed by different pulsars at different frequency bands. Our data show lower jitter and hence more precise DMs at lower  frequencies (than higher frequencies). However, low frequency studies of other pulsars are required to see if this trend is universal. It is very well possible that the trend is opposite for other pulsars and this is a subject of a subsequent analysis.  

\section{Conclusions}
\label{sec:Conclusions}

In this work, we present the first ever jitter measurements of PSR~J0437$-$4715 in the  low frequency region (300 MHz - 500 MHz), using Band 3 wideband observations obtained as a part of the InPTA experiment.
We were able to estimate the  jitter values in both Band 3 and 5 using  Bayesian inference. 
For Band 5, our jitter measurements are in agreement with the previous studies; (\citetalias{Shannon2014} and \citetalias{Parthasarathy2021}).  In this band, we estimate the weighted average of jitter to be  $56.2 \pm 1.6$~ns. Our Band 3 weighted average value for the jitter is  $53.2 \pm 0.9$~ns, which is lower than that in Band 5. 
\bthis{Therefore, we see a positive correlation in the frequency dependence of jitter at low frequencies in the 300-500 MHz range.}
To explore the reason for this, we measured the jitter amplitude in each frequency sub-band also, using sub-banded TOAs from Band 3 observations. The results from the sub-banded data reaffirmed the positive correlation of the jitter amplitude with frequency. Therefore, positive correlation of the  jitter with frequency in our low frequency (Band 3) analysis in tandem with the negative correlation of jitter with frequency in high frequency analysis by \citetalias{Shannon2014} and \citetalias{Parthasarathy2021} suggests the existence of a \textit{turnover} frequency region for the jitter amplitude. This \textit{turnover} region probably lies somewhere between 500 and 700 MHz and this explains why our jitter amplitude of Band 3 is not as large as expected from previous high frequency studies alone. It would be interesting to explore this \textit{turnover} region further and this \textit{turnover} frequency could possibly be determined by the uGMRT Band 4 (550 - 750 MHz) observations, or the next generation, high sensitivity telescopes, such as the SKA.

We have also explored the DM measurements for short duration observations and the effect of jitter on its precision. Using two independent approaches, we analysed 10 second sub-integrations of a high S/N epoch to estimate the error in DM measurements. We found that the values inferred from the observational data alone are in good agreement with the analysis done using quasi-theoretical approach, using the estimated jitter values for this pulsar. We conclude that for high S/N \textit{but} short duration observations, jitter is the dominant source of noise and limits the DM precision which for this pulsar is around $10^{-5}$.

Our interesting results were achieved thanks to the high sensitivity of the uGMRT Band 3 observations.   
Previous studies have shown that lower frequencies suffer more from jitter noise, but our results suggest otherwise. This may, however, not always be the case. For J0437$-$4715, the jitter amplitude in Band 3 is about the same as in 1400 MHz, which shows that observations at low frequency are not severely affected by the jitter noise, and we may be able  obtain more stable TOAs from observations at frequencies below 300 MHz. Also, low frequency jitter studies of other pulsars are needed for making any conclusive statements on the generic jitter behaviour for pulsars on the whole. Further, for a detailed look into the reason for this opposite trend in jitter, single pulse analysis would  be of immense help.
Our future studies will provide single pulse analysis of PSR~J0437$-$4715 and other bright pulsars, as well as comprehensive jitter measurements of the whole InPTA pulsar set, which is important for future timing observation strategies for all PTAs.
%

\begin{acknowledgement}
We acknowledge the GMRT telescope operators for the observations. The GMRT is run by the National Centre for Radio Astrophysics of the Tata Institute of Fundamental Research, India.
\end{acknowledgement}

\paragraph{Funding Statement}

TK is supported by the Terada-Torahiko Fellowship and the JSPS Overseas Challenge Program for Young Researchers.
SD is partially supported by T-641 (DST-ICPS)
BCJ acknowledges support from Raja Ramanna Chair (Track-I) grant from the Department of Atomic Energy, Government of India. BCJ acknowledges support from the Department of Atomic Energy, Government of India, under project number 12-R\&D-TFR-5.02-0700.
KT is partially supported by JSPS KAKENHI Grant Numbers 20H00180, 21H01130, and 21H04467, Bilateral Joint Research Projects of JSPS, and the ISM Cooperative Research Program (2023-ISMCRP-2046).
AmS is supported by CSIR fellowship Grant number 09/1001(12656)/2021-EMR-I and DST-ICPS T-641.
AKP is supported by CSIR fellowship Grant number 09 /0079\\
(15784)/2022-EMR-I.
DD acknowledges the support from the Department of Atomic Energy, Government of India through "Apex Project - Advance Research and Education in Mathematical Sciences at IMSc".
JS acknowledges funding from the South African Research Chairs
Initiative of the Department of Science and Technology and the National 
Research Foundation of South Africa.
SD  acknowledge the support of the Department of Atomic Energy, Government of India, under project identification \# RTI 4002.
YG acknowledges the support from the Department of Atomic Energy, Government of India, under project No. 12-R\&D-TFR-5.02-0700.




\paragraph{Data Availability Statement}

The data underlying this article will be shared on reasonable request to the corresponding author.

\printendnotes

\bibliography{J0437_Jitter.bib}

\appendix

\end{document}